\documentclass[preprint,12pt]{elsarticle}

\usepackage{amsmath,amssymb,amsthm}
\usepackage{url}
\usepackage{subfig}
\usepackage{setspace}

\usepackage{amssymb}

\journal{International Communications in Heat and Mass Transfer }
\usepackage[normalem]{ulem}
\begin{document}

\begin{frontmatter}



\title{Sensitivity of airborne transmission of enveloped viruses to seasonal variation in indoor relative humidity}


\author[inst1]{Alison J. Robey}

\affiliation[inst1]{organization={Center for Environmental Studies, Williams College}}

\author[inst2,inst3]{Laura Fierce\corref{cor1}}
\affiliation[inst2]{organization={Atmospheric Sciences \& Global Change Division, Pacific Northwest National Laboratory}}
\affiliation[inst2]{organization={previously at: Environmental \& Climate Sciences Department, Brookhaven National Laboratory}}


\cortext[cor1]{To whom correspondence should be addressed. E-mail: laura.fierce@pnnl.gov}

\begin{abstract}
In temperate climates, infection rates of enveloped viruses peak during the winter. While these seasonal trends are established in influenza and human coronaviruses, the mechanisms driving the variation remain poorly understood and thus difficult to extend to similar viruses like SARS-CoV-2. In this study, we use the Quadrature-based model of Respiratory Aerosol and Droplets (QuaRAD) to explore the sensitivity of airborne transmission to the seasonal variation in indoor relative humidity across the wide range of relevant conditions, using SARS-CoV-2 as an example. Relative humidity impacts the evaporation rate and equilibrium size of airborne particles, which in turn may impact particle removal rates and virion viability. Across a large ensemble of scenarios, we found that the dry indoor conditions typical of the winter season lead to slower inactivation than in the more humid summer season; in poorly ventilated spaces, this reduction in inactivation rates increases the concentration of active virions, but this effect was important when the susceptible person was farther than 2~m downwind of the infectious person. On the other hand, changes in particle settling velocity with relative humidity did not significantly affect the removal or travel distance of virus-laden scenarios. 
\end{abstract}




\end{frontmatter}


\section{Introduction}
Late autumn to early spring is often referred to not as the winter season, but as the flu season, reflecting the seasonal variation in influenza. Infection rates of influenza surge in the winter \citep{tamerius2011}, similar to other airborne viruses sharing its enveloped structure and upper respiratory infection site \citep{tamerius2013, yang2011}. One potential cause of this trend is the seasonal variation of environmental conditions. While outdoor temperature and humidity levels vary dramatically throughout the year, humans spend around 90\% of their time indoors \citep{klepeis2001, schweizer2007} where temperatures remain stable but relative humidity (RH) displays consistent summer highs and winter lows \citep{marr2019, tamerius2013_2}. Several mechanisms pertinent to airborne transmission of viruses are sensitive to this variability in RH, such as faster inactivation of enveloped virions at extreme RH levels \citep{yang2011, marr2019} and shorter travel distance of mid-sized particles at high RH levels \citep{wei2015}. However, it is unclear whether the impact of RH on either process is important enough to contribute to virus seasonality.

In the context of familiar airborne diseases like the flu or the winter common colds (caused by human coronaviruses (HCoVs) 229E, HKU1, NL63, and OC43) \citep{park2020, stefanea2020}, understanding this relationship between RH and the seasonality of infection rates is a retrospective question; hundreds of years of evidence have clearly established their seasonality \citep{fisman2007}, and current research helps clarify why, not if, seasonal cycles have occurred. In the case of emerging viruses like SARS-CoV-2, however, the question is preemptive: will transmission vary seasonally in the future? The first year of an outbreak cannot answer this question \citep{baker2020, carlson2020}, but establishing whether novel viruses are likely to follow seasonal trends is critical to preparing adequate mitigation strategies \citep{fisman2007, myers2000, rivers2018}.

Determining the impacts of RH on airborne transmission is vital to progressing our understanding of the seasonal dynamics of any enveloped virus. However, quantifying those impacts requires a model holistic enough to account for the series of processes governing airborne transmission and efficient enough to span the tremendous variability in relevant conditions. Most models complex enough to provide a detailed representation of short-range and long-range airborne transmission \citep[e.g.][]{beghein2005, choi2012} are too computationally expensive to represent the large uncertainty in indoor conditions, particle properties, and physiological parameters. To obtain a mechanistic understanding of the impact of RH on airborne transmission, while also accounting for the large uncertainties in model parameters, we performed thousands of simulations using the new Quadrature-based model of Respiratory Aerosol and Droplets (QuaRAD) \citep{fierce2021}. We use the transmission of SARS-CoV-2 as an example due to its current relevance; however, conclusions are also relevant to other prominent winter viruses that share its enveloped virion structure \citep{biryukov2020} and ability to transmit through the airborne route \citep{samet2021}.

In this paper, we first describe QuARAD (Section~\ref{sec:methods}), emphasizing specific details about the interaction of RH with evaporation (Section~\ref{subsec:evap}) and viral inactivation (Section~\ref{subsec:inact}). We then examine the impacts of typical summer and winter RH levels on exposure in a single baseline scenario (Section~\ref{subsec:scenario}) and across a large ensemble of scenarios (Section~\ref{subsec:ensemble}). Finally, we discuss the insights gained on the seasonality of airborne enveloped viruses and the utility of different mitigation strategies (Section~\ref{sec:discuss}).

\section{Model Description}\label{sec:methods}
We assessed the impact of typical indoor RH levels on virion fate and transport using QuaRAD, a broad overview of which is given below. The full model description, including specific parameter distributions, can be found in \cite{fierce2021}. The model approximates the size distribution of and viral load within expired particles using numerical quadrature. Such quadrature-based moment methods have been shown to accurately represent integrals over aerosol distributions with only a small number of particles \citep{mcgraw1997, fierce2017}.

QuaRAD models the evolution of particles expired by an infectious individual and the subsequent exposure of a susceptible individual. The particle size distribution is represented with three superimposed lognormal modes, based on the measurements of speech emissions detailed in \cite{johnson2011}. We estimate the viral loads associated with each particle using these size distribution measurements in combination with Gesundheit II and quantitative RT-PCR measurements of influenza virion emission on fine and coarse particles \citep{milton2013,leung2020}. We assume that influenza is a reasonable model for viral emissions of SARS-CoV-2 \citep{jacot2020}. In QuaRAD, the size distribution of respiratory particles and its evolution is represented using only six weighted particles. 

For each quadrature point, we simulate the evaporation (Section~\ref{subsec:evap}) and the subsequent virion inactivation (Section~\ref{subsec:inact}). Particle dispersion within the expiratory jet of an infectious person is modeled using a Gaussian puff model \citep{drivas1996} in combination with a buoyant turbulent jet model \cite{lee2003}, following the approach of \cite{wei2015}.  We then sum over the quadrature points to approximate the integral of the size distribution inhaled by the susceptible individual. Input parameters are sampled from the realistic probability distributions detailed in \cite{fierce2021}; the baseline scenario discussed in Section~\ref{subsec:scenario} was simulated using the median of the probability density function from which the ensemble inputs were sampled.

In this study, we analyzed the continuous stream of aerosol particles and droplets expelled when an infectious person speaks and the subsequent exposure of a susceptible individual, which we quantified as the number of virions inhaled per second $\dot{N}_{\text{inhale}}$. $\dot{N}_{\text{inhale}}$ was computed using the concentration of virions present at the mouth of the susceptible person, their average number of breaths per minute, and the average volume of air inhaled per breath. We quantify the difference in active virion inhalation between RH cases as described in Section~\ref{subsec:diff}.

\subsection{Evaporation}\label{subsec:evap}
Upon expiration, aerosol particles and droplets evaporate to an equilibrium size that depends on their composition and the environmental conditions. Following the approach of \cite{wei2015} and \cite{kukkonen1989} and implemented as detailed in \cite{fierce2021}, we model evaporation by solving a coupled pair of ordinary differential equations representing the mass and heat transfer:
\begin{align}\label{eqn:evaporation_m}
&\frac{dm_{\text{p}}}{dt}=\frac{2\pi pD_{\text{p}}M_{\text{w}}D_{\infty}C_{\text{T}}\text{Sh}}{RT_{\infty}}\ln\bigg(\frac{p-p_{\text{v,p}}}{p-p_{\text{v},\infty}}\bigg)
\end{align}
\begin{align}\label{eqn:evaporation_T}
&\frac{dT_{\text{p}}}{dt}=\frac{1}{m_{\text{p}}C_{\text{p}}}\bigg(\pi D_{\text{p}}^2k_g\frac{T_{\infty} - T_{\text{p}}}{0.5D_{\text{p}}}\text{Nu}-L_{\text{v}}\frac{dm_{\text{p}}}{dt}\bigg).
\end{align}
Equation~\ref{eqn:evaporation_m} describes the temporal evolution of the particle mass $m_{\text{p}}$ as a function of time $t$, ambient pressure $p$, particle diameter $D_{\text{p}}$, molecular weight of water $M_{\text{w}}$, diffusivity of water in air $D_{\infty}$, correction factor $C_{\text{T}}$, Sherwood number Sh, gas constant $R$, surface vapor pressure $p_{\text{v,p}}$, and ambient vapor pressure $p_{\text{v},\infty}$. Equation~\ref{eqn:evaporation_T} describes the evolution of the particle temperature $T_{\text{p}}$ as a function of the particle's specific heat $C_{\text{p}}$, the air thermal conductivity $k_{\text{g}}$, the background temperature $T_{\infty}$, the Nusselt number Nu, the latent heat of vaporization $L_{\text{v}}$, and Equation~\ref{eqn:evaporation_m}. These parameters are provided in \cite{fierce2021}. The model represents convection-induced evaporation enhancements. We assume that particles are spherical and contain 1\%--9\% non-water aerosol components by volume, following the measurements of \cite{vejerano2018}. The vapor pressure over an aqueous droplet is computed from the $\kappa$-K\"{o}hler model \citep{petters2007}.

\begin{figure}
    \centering
    \includegraphics[width=4.in]{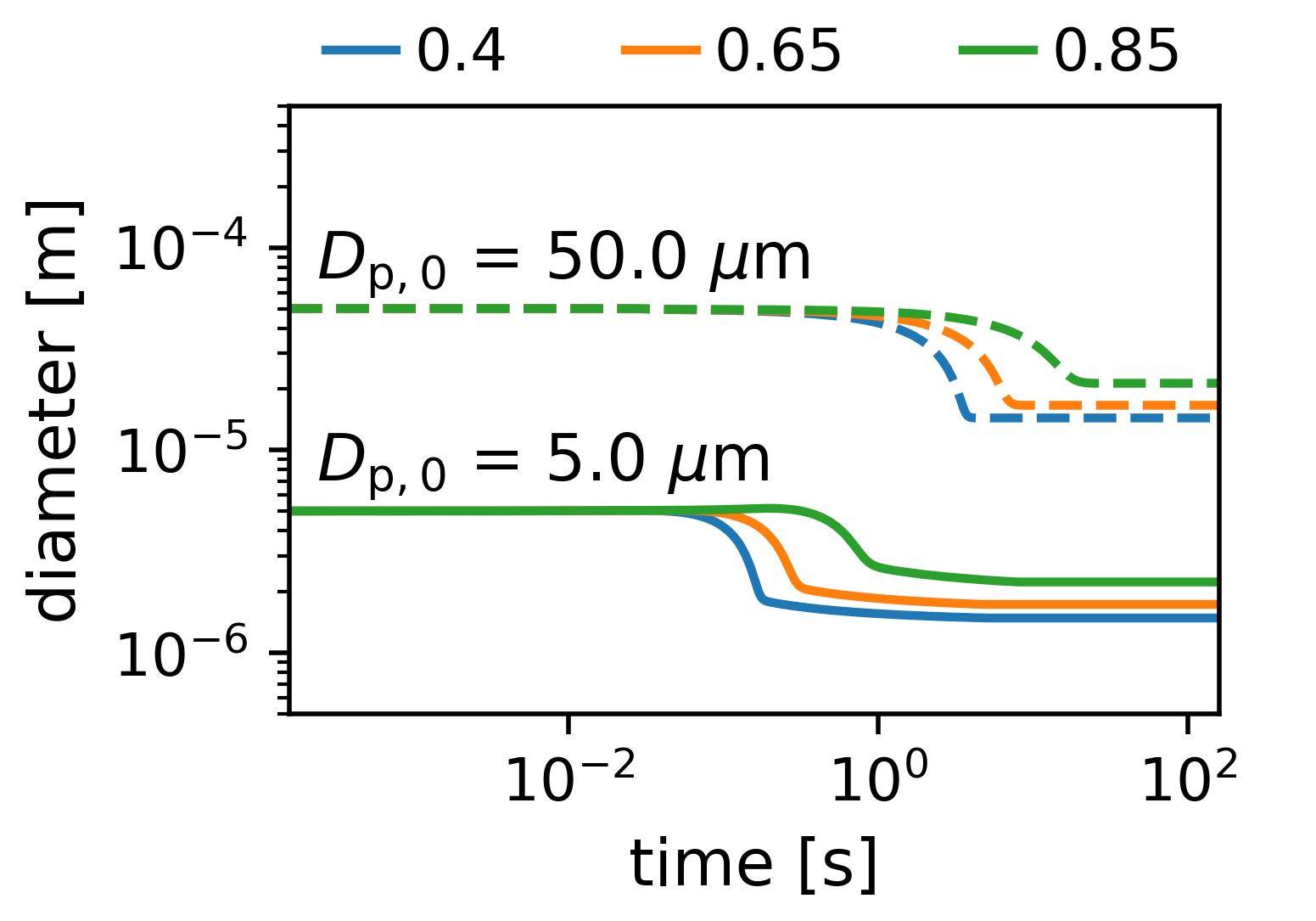}
    \caption{The evolution of particle diameter $D_{\text{p}}$ with respect to time due to evaporation at different RH levels.}
    \label{fig:evap}
\end{figure}

Seasonal changes in RH modify $p_{\text{v},\infty}$, such that droplets evaporate more slowly with increasing RH, as shown in Figure~\ref{fig:evap}. RH is the ratio between $p_{\text{v},\infty}$ and the saturation vapor pressure, which depends on temperature. Although QuaRAD is able to represent the variation in temperature and RH within the turbulent jet of an infectious person, we found that results were insensitive to this variation. Therefore, we assume that expired particles encounter the background temperature and humidity as soon as they are expired. 

\subsection{Viral Inactivation}\label{subsec:inact}
\begin{figure}[h!]
    \centering
    \includegraphics[width=3.in]{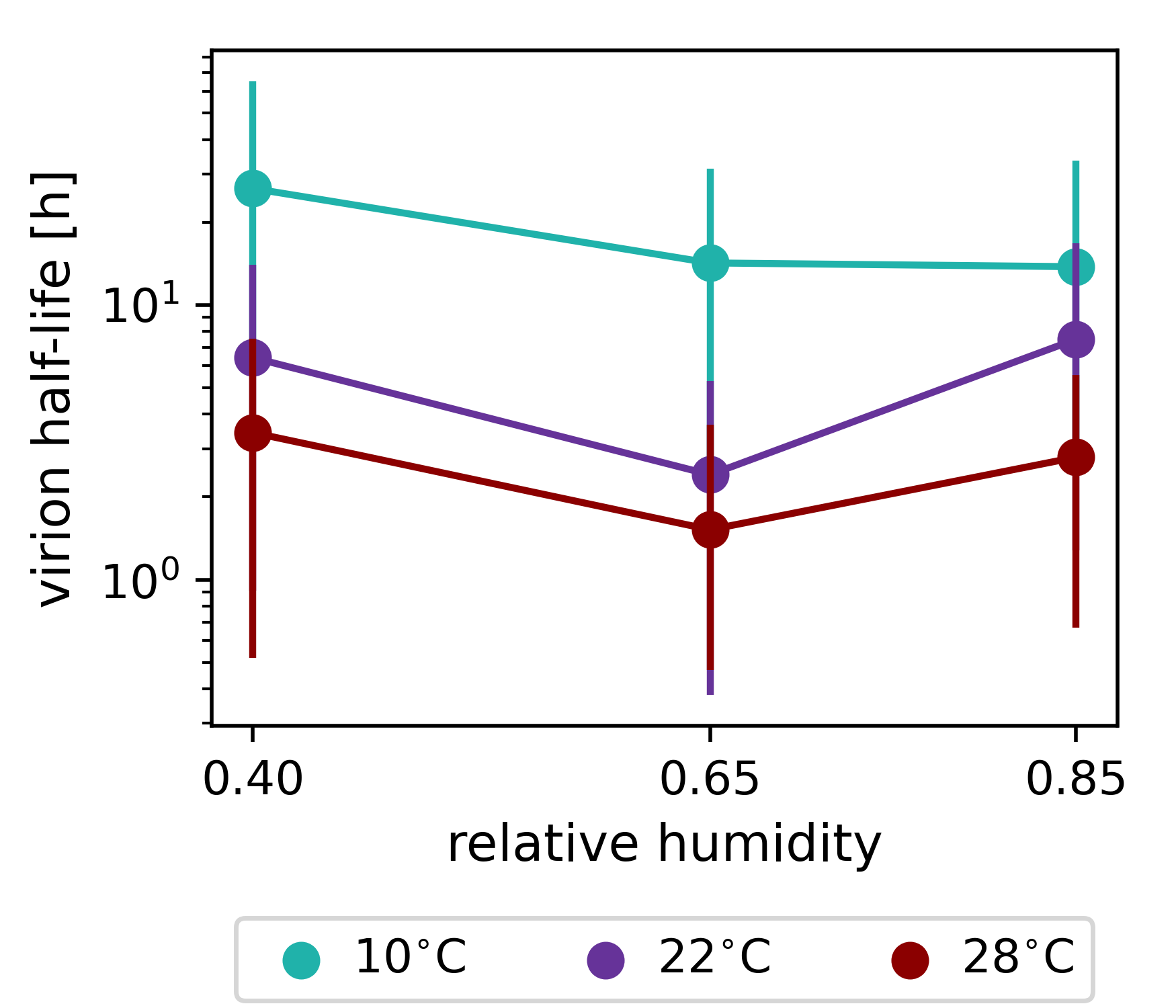}
    \caption{Measured SARS-CoV-2 virion half-lives under different conditions adapted from \cite{morris2020}. We focused on the half-lives of 6.43~h at RH~$=40\%$ and 2.41~h at RH~$=65\%$ measured at room temperature (shown in purple), corresponding to inactivation rates of 0.108~h$^{-1}$ and 0.288~h$^{-1}$, respectively.}
    \label{fig:morris}
\end{figure}

Like other enveloped viruses \citep{marr2019}, SARS-CoV-2 virions exhibit exponential decay \citep{morris2020} with comparable rates on surfaces and while airborne \citep{biryukov2020, van2020}. The inactivation of a virion eliminates its capability of infecting a cell and is analogous to the death of an organism. The trends of inactivation due to temperature and RH are established, though the exact inactivation rates remain poorly constrained; since our focus is on the indoor environment, we do not consider the accelerated inactivation rates caused by UV light \citep{carlson2020}. As with influenza \citep{marr2019, yang2012} and other coronaviruses \citep{casanova2010, songer1967}, SARS-CoV-2 decays more rapidly at high temperatures than low temperatures and shows a U-shaped dependence on RH, as seen in Figure~\ref{fig:morris}. At room temperature, for example, virion half-lives are longer than 6 hours for RH~$<45\%$ or RH~$>75\%$ but are around 2.5 hours for RH~$\approx65\%$ \citep{morris2020}.

Several mechanisms may cause the temperature dependence of viral inactivation, including thermal denaturation \citep{marr2019, morris2020, woese1960} and decreased lipid ordering in viral envelopes at high temperatures \citep{marr2019, polozov2008}. The RH dependence of viral inactivation is likely caused by RH's influence over solute concentrations in infectious particles. Solutes are thought to function as reactants in virion inactivation, and the U-shaped humidity dependence is thus theorized to result from two ways in which the interior solute concentrations are reduced \citep{morris2020, lin2019}. At low RH levels, evaporation causes solutes in expired particles to effloresce, separating from the aqueous solution as solids on the particle surface. If the RH within the indoor space is above a threshold, known as the efflorescence relative humidity (ERH), solutes within the aqueous particle will remain in the solution. In this regime, the solute concentration decreases with increasing RH because, as long as the RH remains above the ERH, the concentration of solutes will remain more dilute the more water is removed through evaporation. For this reason, inactivation rates are fastest just above the ERH, when solutes are in solution and are also highly concentrated.

In this model, we used the measured distributions of SARS-CoV-2 half-lives $h$ in a cell culture medium at three temperatures (283.15 K, 295.15 K, 300.15 K) and RH levels (40\%, 65\%, 85\%) from \cite{morris2020} to estimate the distribution of viral decay rates $k$ given in Figure~\ref{fig:morris}. Though we recognize that ERH varies with particle size and composition, we followed \cite{morris2020} in assuming an ERH of $45\%$ across particles. These values were chosen due to the explicit variation in environmental conditions but are comparable to those obtained in \cite{biryukov2020}, \cite{van2020}, and \cite{kwon2020}. Given the focus of this paper on seasonal indoor environments, we analyzed the measurements taken at T~$=295.15$~K, corresponding to an average indoor temperature of 22$^{\circ}$C, and at RH~$=40\%$ or RH~$=65\%$.

\subsection{Quantifying differences caused by RH}\label{subsec:diff}
We quantify the effect of RH on virion exposure as the relative difference between virion inhalation rates, $\dot{N}_{\text{inhale}}$, at two RH levels:
\begin{equation}\label{eqn:rel_diff}
    \Delta_{\text{r}} = \frac{\dot{N}_{\text{inhale, RH}=40\%} - \dot{N}_{\text{inhale, RH}=65\%}}{\dot{N}_{\text{inhale, RH}=65\%}}
\end{equation}
Positive values of the relative difference $\Delta_{\text{r}}$ indicate an increase in the inhalation rate of active virions with a decrease in RH from 65\% to 40\%. These RH levels match the conditions under which \cite{morris2020} measured virion half-lives and approximately correspond to the median indoor RH of $62\%$ in the summer and $42\%$ in the winter measured by \cite{nguyen2014}. Susceptible individuals are assumed to inhale at a rate of $14$ breaths per minute and a volume of $4.69\times10^{-4}$ m$^3$ per breath in the base case and at 12 to 20 breaths per minute and $3.75\times10^{-4}$ to $6.25\times10^{-4}$ m$^3$ per breath across the ensemble \citep{fierce2021}.

\section{Results}\label{sec:results}
In this section, we present the sensitivity of virion inhalation rate to RH as a function of downwind distance from an infectious individual. Before describing the effect of RH on airborne transmission across an ensemble of simulations in Section~\ref{subsec:ensemble}, we first show how RH impacts key processes in an example scenario in Section~\ref{subsec:scenario}.

\subsection{Sensitivity of virion exposure to RH in a single scenario}\label{subsec:scenario}
\begin{figure}[h!]
    \centering
    \includegraphics[width=2.5in]{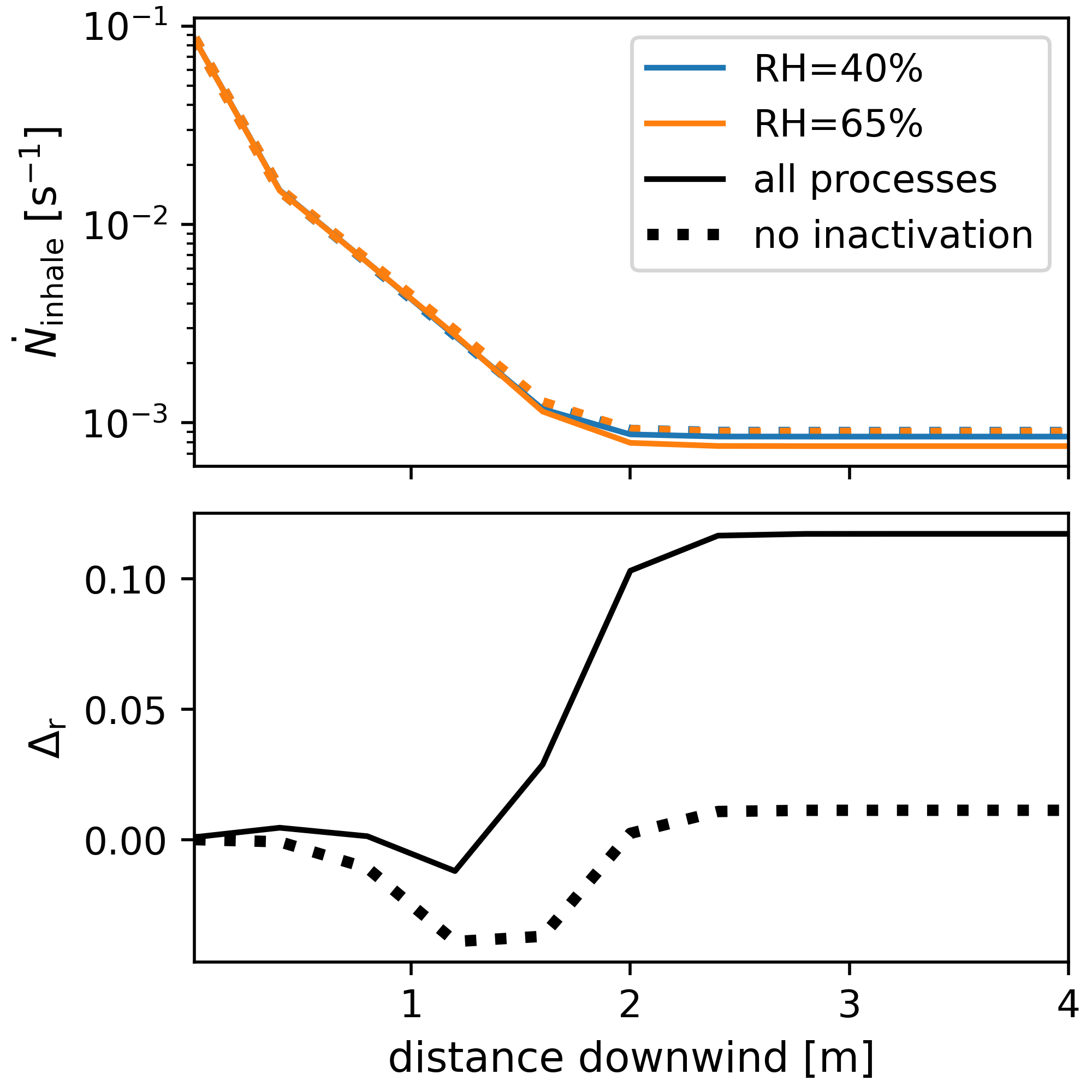}
    \caption{The (A) virion inhalation rate $\dot{N}_{\text{inhale}}$ and (B) relative difference between RH levels $\Delta_{\text{r}}$ computed with Equation~\ref{eqn:rel_diff} in the baseline scenario with respect to downwind distance after 1.5 hours.}
    \label{fig:compare}
\end{figure}

In the baseline scenario, we found that the sensitivity of $\dot{N}_{\text{inhale}}$ to RH depends on the distance between the infectious and the susceptible person (Figure~\ref{fig:compare}). When directly downwind of an infectious person, $\dot{N}_{\text{inhale}}$ is insensitive to RH, whereas we find a 10\% increase in $\dot{N}_{\text{inhale}}$ with a decrease in RH from 65\% to 40\% (solid line in Figure~\ref{fig:compare}B). When inactivation is neglected, $\Delta_{\text{r}} \approx 0$ both near and far from the infectious person (dotted line in Figure~\ref{fig:compare}B).

If a susceptible person is within the expiratory jet of an infectious person, nearly all virions they inhale would have been expelled from an infectious person just seconds earlier. On the other hand, far from an infectious person, a large portion of virions may be contained in particles that were expelled hours earlier. For this reason, $\dot{N}_{\text{inhale}}$ far from an infectious person depends strongly on the RH-dependent inactivation rates ($\Delta_{\text{r}}\approx0.1$), whereas near-field exposure to $\dot{N}_{\text{inhale}}$ is insensitive to changes in inactivation rates ($\Delta_{\text{r}}\approx0$).

\subsection{Factors governing sensitivity of virion exposure to RH across ensemble}\label{subsec:ensemble}
\begin{figure}[!ht]
    \centering
    \includegraphics[width=4.5in]{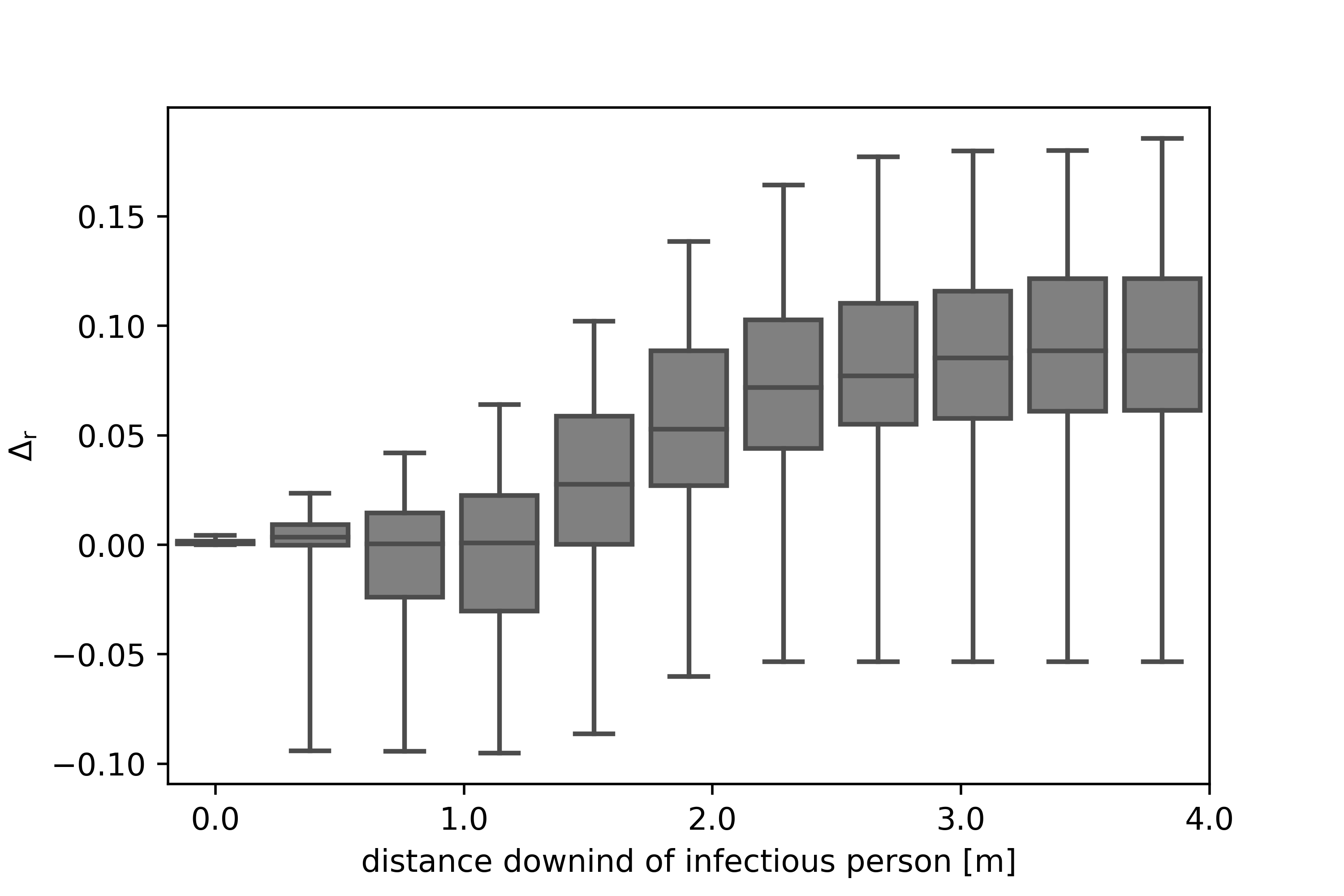}
    \caption{The median (solid line), quartiles (boxes), and 90\% confidence interval (whiskers) of the relative difference $\Delta_{\text{r}}$ between RH levels computed with Equation~\ref{eqn:rel_diff} at each downwind distance for 1000 scenarios after 4 hours with ACH~$=1.5 \pm 1.2$.}
    \label{fig:overall}
\end{figure}

To quantify the impact of RH across the wide range of conditions expected in indoor spaces, we performed an ensemble of 1000 scenarios using the distributions in input parameters described in \cite{fierce2021}. As in the base case, $\dot{N}_{\text{inhale}}$ was insensitive to RH when directly downwind of an infectious person. Beyond distances of approximately 2~m, $\dot{N}_{\text{inhale}}$ was higher at a typical winter RH of 40\% than at a typical summer RH of 65\%. The mean $\Delta_{r}$ (Figure~\ref{fig:overall}) follows the same trend as the base case. We again found that the sensitivity of $\dot{N}_{\text{inhale}}$ to RH was driven by differences in virion inactivation and not by differences in particle settling.
 
\begin{figure}
    \centering
    \includegraphics[width=5.in]{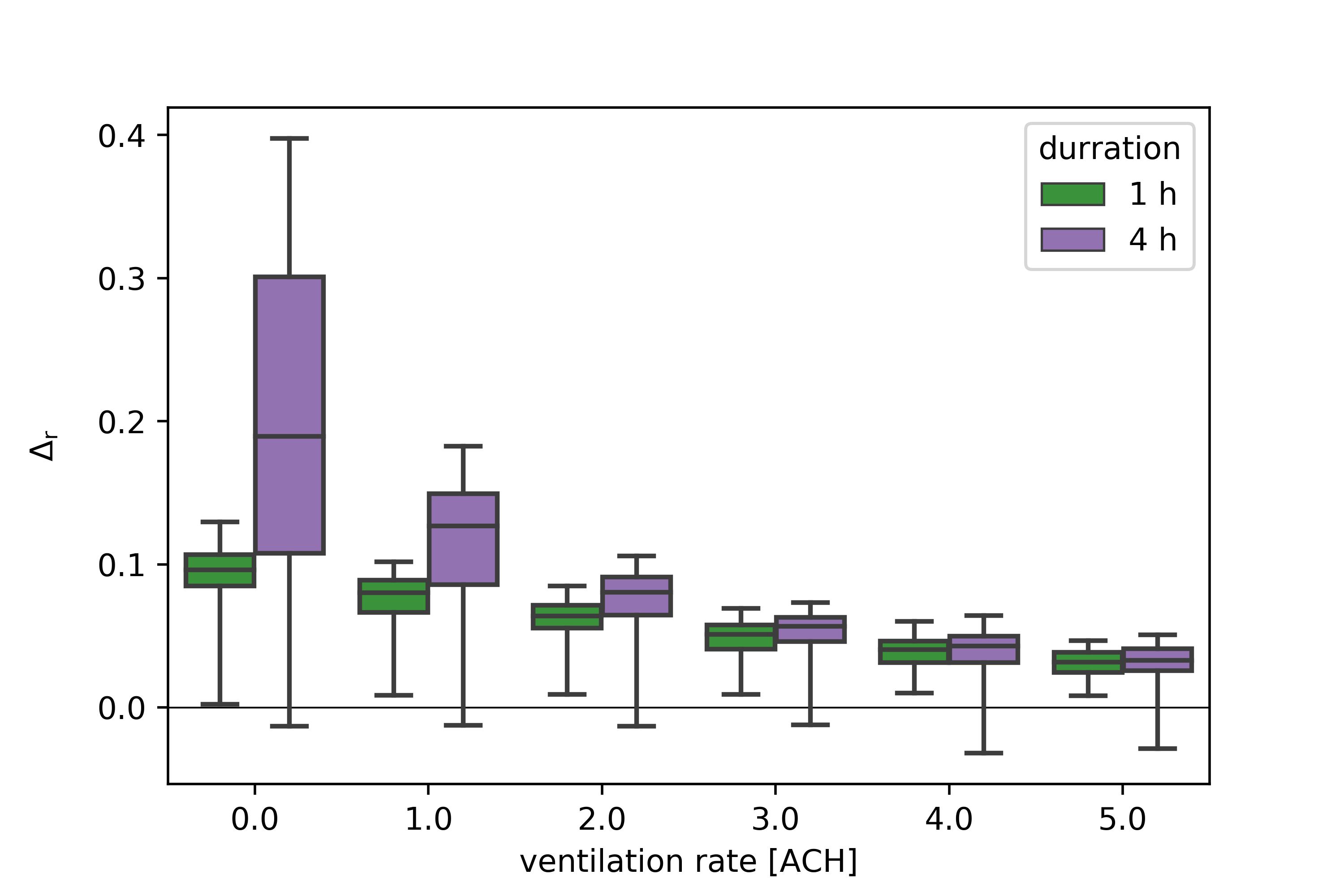}
    \caption{Comparison between median (solid line), quartiles (boxes), and 90\% confidence interval (whiskers) in $\Delta_{\text{r}}$ after 1~h and 4~h at each ventilation rate for individuals standing 4~m apart.}
    \label{fig:ACH}
\end{figure}

Since RH significantly affects transmission only when the susceptible person is far from the infectious person, we explored the sensitivity of $\dot{N}_{\text{inhale}}$ to conditions controlling far-field virion concentrations: the duration of the encounter and the ventilation rate of the room (expressed here in air changes per hour, or ACH). If the infectious person and the susceptible person are in a poorly ventilated space, the virion concentrations throughout the room increase over time. Under these conditions, virion exposure is highly sensitive to inactivation rate, as shown by the high $\Delta_{\text{r}}$ values at low ventilation rates in Figure~\ref{fig:ACH}. Virion concentrations also increase as the duration of the encounter increases. Consequently, longer interactions correspond with higher exposure and a higher $\Delta_{\text{r}}$ (comparison between $\Delta_{\text{r}}$ in a 1-hour interaction (green) and in a 4-hour interaction (purple) in Figure~\ref{fig:ACH}).

We also found that high ventilation rates reduce the dependence of $\Delta_{\text{r}}$ on the length of the encounter. For example, when the ventilation rate is less than $1.0$~ACH, the median $\Delta_{\text{r}}$ after four hours is nearly twice the median $\Delta_{\text{r}}$ after one hour, but when the ventilation rate is greater than $3.0$~ACH, the median $\Delta_{\text{r}}$ values after four hours and after one hour are approximately equal.

\section{Discussion}\label{sec:discuss}

\subsection{Comparison between mechanisms}
Our process analysis revealed that the greater virion exposure at RH~$=40\%$ than at RH~$=65\%$ was caused by differences in virion inactivation rate. In indoor spaces, SARS-CoV-2 virions have long half-lives ($h=6.4 \pm 0.03$~hours at RH~$=40\%$ and $h=2.4 \pm 0.04$~hours at RH~$=65\%$), so nearly all virions in freshly expired particles are active, regardless of the background RH. Whereas virion exposure directly downwind of an infectious person is driven by virions in freshly expired particles, virion exposure far from an infectious person may be strongly influenced by particles that remain suspended for hours.

Whereas changes in inactivation rates with RH led to seasonal differences in virion survival within respiratory particles, removal rates and travel distances of the respiratory particles themselves were insensitive to seasonal changes in RH. RH does affect the evaporation timescale and equilibrium size of expired particles (see Figure~\ref{fig:evap}), but the impact of those differences on travel distance is limited by the expired size distribution. Close to the infectious individual, most expired particles are small enough to travel with the flow of the respiratory jet at their initial diameter, during evaporation, and at their equilibrium diameter; most of the other expired particles are large enough to quickly settle out of the respiratory jet at their initial diameter, during evaporation, and at their equilibrium diameter. Far from the infectious individual, the equilibrium sizes of most particles at RH~$=40\%$ and RH~$=65\%$ are too similar to cause major differences in particle travel or residence time.
The plume does contain mid-sized particles whose spread and settling velocity are sensitive to RH, but these particles make up only a tiny proportion of those expired during speech and consequently have very little effect on $\dot{N}_{\text{inhale}}$. While this finding does not contradict earlier findings of RH sensitivities in the spread of specific particles, it does show that those differences have very little impact on transmission risk when integrating over a population of polydisperse particles.

\subsection{Sensitivity to timescale and ventilation rate}
Analysis across the ensemble revealed that at 4~m of distance, RH has a much greater impact on virion exposure over the course of a long encounter than over the course of a short one. Far-field concentrations increase over time, such that inhaling air in the far-field after only a few minutes of emissions results in a much lower $\dot{N}_{\text{inhale}}$ than after an hour of emissions; near-field concentrations do not show this time dependence because, in close proximity, $\dot{N}_{\text{inhale}}$ is controlled by the freshly expired virions in the expiratory jet instead of the virions building up over time throughout the room. Because longer timescales result in greater virion concentrations than shorter timescales and the decay rate of far-field virion concentrations depend on RH, longer timescales thus allow the decay rate to have a greater impact on $\dot{N}_{\text{inhale}}$, resulting in an elevated $\Delta_{\text{r}}$.

Far-field concentrations are also affected by the ventilation rate of the room. Ventilation removes particles from the air, so when the ACH is low, far-field particle concentrations build up over time. When particles linger for a long time, the difference in inactivation rate between RH scenarios results in a high $\Delta_{\text{r}}$. On the other hand, when the ACH is high, far-field particle concentrations are reduced more quickly. The impact of RH on exposure thus decreases as the ACH increases, resulting in a small $\Delta_{\text{r}}$ in well-ventilated spaces. We thus find that the impact of RH on far-field exposure over long timescales can be meaningfully reduced by increasing the ventilation rate.

\subsection{Study limitations}
While we found that typical winter and summer indoor RH levels indeed impact virion exposure, it is very likely that other factors not accounted for in this model are also in play. For example, immune response capabilities of susceptible individuals may be impaired by seasonal deficiencies in vitamin D \citep{cannell2006} or RH-mediated reductions in mucociliary clearance \citep{pieterse2018}, each of which may increase the probability of winter infections. Frequency of exposure may also be determined by seasonal differences in human behavior, including the school calendar and the higher frequency of indoor social events in colder seasons \citep{earn2012, vouriot2021}. We also note that QuaRAD accounts solely for seasonal differences in airborne transmission; while this is likely the dominant transmission route of SARS-CoV-2 \citep{greenhalgh2021} and a prominent one for other discussed viruses, differences in droplet or fomite transmission may independently impact seasonality.

\section{Conclusion}
In this paper, we showed that a low RH increases airborne virion exposure by slowing the virion inactivation rate, not by increasing the travel distance or residence time of respiratory particles. In poorly ventilated spaces and encounters lasting four hours, we found that lowering the RH from 65\% to 40\% led to a 20\% decrease in the median exposure to active virions when the susceptible individual is at least 2~m from the infectious individual. Close-range exposure was insensitive to this seasonal variation in indoor RH. Whereas previous studies \citep[e.g.][]{wei2015} have shown that changes in RH can significantly impact the dispersion of particles in specific size ranges, we found that differences in evaporation speed and equilibrium size have a trivial impact on the overall dispersion due to the tiny fraction of the plume those particles make up. The timescale at which virions become inactive, on the other hand, is highly sensitive to RH and thus strongly impacts the inhalation rate of active virions. This finding highlights the importance of measurements that constrain the relationship between virion inactivation and environmental properties \citep[e.g.][]{morris2020}.

Though we focused our analysis on SARS-CoV-2, we expect airborne exposure to other enveloped viruses to follow the same trends. We note that the sensitivity of virion inhalation rate to RH will be more pronounced for viruses with shorter half-lives and, furthermore, that while we exclusively studied the \emph{rate} of virions inhaled, it is the absolute \emph{number} of virions inhaled that determines the probability of initial infection. As such, small sensitivities in inhalation rate to RH may, over time, result in larger sensitivities in the actual probability of infection to RH. The observed sensitivity of virion exposure to RH is reduced with higher ventilation and shorter exposure times, suggesting high ventilation and shorter interactions are critical to limiting winter transmission risk even when following safe social distancing guidelines. In settings where short exposure times are impossible or where multiple groups of people transition between the same room, such as classrooms, hospitals, factories, or prisons, employing other protections like effective mask wearing and high ventilation are critical to limiting transmission risks.

\section*{Data Availability}
The QuaRAD source code, input files, and processing script are available for download at: \url{https://github.com/lfierce2/QuaRAD/}. Simulation ensembles were created using latin hypercube sampling with pyDOE: \url{https://pythonhosted.org/pyDOE/}.

\section*{Acknowledgements}
This research was supported by the DOE Office of Science through the National Virtual Biotechnology Laboratory, a consortium of DOE national laboratories focused on response to COVID-19, with funding provided by the Coronavirus CARES Act. This project was supported in part by the U.S. Department of Energy through the Office of Science, Office of Workforce Development for Teachers and Scientists (WDTS) under the Science Undergraduate Laboratory Internships Program (SULI) and the Environmental and Climate Sciences Department under the BNL Supplemental Undergraduate Research Program (SURP). The quadrature-based model was originally developed for simulation of atmospheric aerosol with support from the DOE Atmospheric System Research program at Brookhaven National Laboratory, a multiprogram national laboratory supported by DOE Contract DE-SC0012704. The authors would like to thank Robert McGraw for helpful discussion and his insights on particle chemistry.



 \bibliographystyle{elsarticle-num} 
 \bibliography{rh_3}

\begin{thebibliography}{10}
\expandafter\ifx\csname url\endcsname\relax
  \def\url#1{\texttt{#1}}\fi
\expandafter\ifx\csname urlprefix\endcsname\relax\def\urlprefix{URL }\fi
\expandafter\ifx\csname href\endcsname\relax
  \def\href#1#2{#2} \def\path#1{#1}\fi

\bibitem{tamerius2011}
J.~Tamerius, M.~I. Nelson, S.~Z. Zhou, C.~Viboud, M.~A. Miller, W.~J. Alonso,
  Global influenza seasonality: reconciling patterns across temperate and
  tropical regions, Environmental health perspectives 119~(4) (2011) 439--445.

\bibitem{tamerius2013}
J.~D. Tamerius, J.~Shaman, W.~J. Alonso, K.~Bloom-Feshbach, C.~K. Uejio,
  A.~Comrie, C.~Viboud, Environmental predictors of seasonal influenza
  epidemics across temperate and tropical climates, PLoS Pathog 9~(3) (2013)
  e1003194.

\bibitem{yang2011}
W.~Yang, L.~C. Marr, Dynamics of airborne influenza a viruses indoors and
  dependence on humidity, PloS one 6~(6) (2011) e21481.

\bibitem{klepeis2001}
N.~E. Klepeis, W.~C. Nelson, W.~R. Ott, J.~P. Robinson, A.~M. Tsang,
  P.~Switzer, J.~V. Behar, S.~C. Hern, W.~H. Engelmann, The national human
  activity pattern survey (nhaps): a resource for assessing exposure to
  environmental pollutants, Journal of Exposure Science \& Environmental
  Epidemiology 11~(3) (2001) 231--252.

\bibitem{schweizer2007}
C.~Schweizer, R.~D. Edwards, L.~Bayer-Oglesby, W.~J. Gauderman, V.~Ilacqua,
  M.~J. Jantunen, H.~K. Lai, M.~Nieuwenhuijsen, N.~K{\"u}nzli, Indoor
  time--microenvironment--activity patterns in seven regions of europe, Journal
  of exposure science \& environmental epidemiology 17~(2) (2007) 170--181.

\bibitem{marr2019}
L.~C. Marr, J.~W. Tang, J.~Van~Mullekom, S.~S. Lakdawala, Mechanistic insights
  into the effect of humidity on airborne influenza virus survival,
  transmission and incidence, Journal of the Royal Society Interface 16~(150)
  (2019) 20180298.

\bibitem{tamerius2013_2}
J.~Tamerius, M.~Perzanowski, L.~Acosta, J.~Jacobson, I.~Goldstein, J.~Quinn,
  A.~Rundle, J.~Shaman, Socioeconomic and outdoor meteorological determinants
  of indoor temperature and humidity in new york city dwellings, Weather,
  Climate, and Society 5~(2) (2013) 168--179.

\bibitem{wei2015}
J.~Wei, Y.~Li, Enhanced spread of expiratory droplets by turbulence in a cough
  jet, Building and Environment 93 (2015) 86--96.

\bibitem{park2020}
S.~Park, Y.~Lee, I.~C. Michelow, Y.~J. Choe, Global seasonality of human
  coronaviruses: a systematic review, Open forum infectious diseases 7~(11)
  (2020) ofaa443.

\bibitem{stefanea2020}
R.~L. Stefanea, M.~J. Binnicker, A.~S. Thomas, R.~Patel, Seasonality of
  coronavirus 229e, hku1, nl63 and oc43 from 2014--2020, in: Mayo Clinic
  Proceedings, Elsevier, 2020, pp. 1--8.

\bibitem{fisman2007}
D.~N. Fisman, Seasonality of infectious diseases, Annu. Rev. Public Health 28
  (2007) 127--143.

\bibitem{baker2020}
R.~E. Baker, W.~Yang, G.~A. Vecchi, C.~J.~E. Metcalf, B.~T. Grenfell,
  Susceptible supply limits the role of climate in the early sars-cov-2
  pandemic, Science 369~(6501) (2020) 315--319.

\bibitem{carlson2020}
C.~J. Carlson, A.~C. Gomez, S.~Bansal, S.~J. Ryan, Misconceptions about weather
  and seasonality must not misguide covid-19 response, Nature Communications
  11~(1) (2020) 1--4.

\bibitem{myers2000}
M.~F. Myers, D.~Rogers, J.~Cox, A.~Flahault, S.~I. Hay, Forecasting disease
  risk for increased epidemic preparedness in public health, Advances in
  parasitology 47 (2000) 309--330.

\bibitem{rivers2018}
C.~M. Rivers, S.~V. Scarpino, Modelling the trajectory of disease outbreaks
  works, Nature 559~(7715) (2018) 477--477.

\bibitem{beghein2005}
C.~Beghein, Y.~Jiang, Q.~Y. Chen, Using large eddy simulation to study particle
  motions in a room, Indoor air 15~(4) (2005) 281--290.

\bibitem{choi2012}
J.-I. Choi, J.~R. Edwards, Large-eddy simulation of human-induced contaminant
  transport in room compartments, Indoor air 22~(1) (2012) 77--87.

\bibitem{fierce2021}
L.~Fierce, A.~Robey, C.~Hamilton, Simulating near-field enhancement in
  transmission of airborne viruses with a quadrature-based model,
  arXiv:2104.01219 [physics.med-ph] (2021).

\bibitem{biryukov2020}
J.~Biryukov, J.~A. Boydston, R.~A. Dunning, J.~J. Yeager, S.~Wood, A.~L. Reese,
  A.~Ferris, D.~Miller, W.~Weaver, N.~E. Zeitouni, et~al., Increasing
  temperature and relative humidity accelerates inactivation of sars-cov-2 on
  surfaces, MSphere 5~(4) (2020).

\bibitem{samet2021}
J.~M. Samet, K.~Prather, G.~Benjamin, S.~Lakdawala, J.-M. Lowe, A.~Reingold,
  J.~Volckens, L.~C. Marr, Airborne transmission of severe acute respiratory
  syndrome coronavirus 2 (sars-cov-2): What we know, Clinical Infectious
  Diseases (2021).

\bibitem{mcgraw1997}
R.~McGraw, Description of aerosol dynamics by the quadrature method of moments,
  Aerosol Science and Technology 27~(2) (1997) 255--265.

\bibitem{fierce2017}
L.~Fierce, R.~L. McGraw, Multivariate quadrature for representing cloud
  condensation nuclei activity of aerosol populations, Journal of Geophysical
  Research: Atmospheres 122~(18) (2017) 9867--9878.

\bibitem{johnson2011}
G.~Johnson, L.~Morawska, Z.~Ristovski, M.~Hargreaves, K.~Mengersen, C.~Y.~H.
  Chao, M.~Wan, Y.~Li, X.~Xie, D.~Katoshevski, et~al., Modality of human
  expired aerosol size distributions, Journal of Aerosol Science 42~(12) (2011)
  839--851.

\bibitem{milton2013}
D.~K. Milton, M.~P. Fabian, B.~J. Cowling, M.~L. Grantham, J.~J. McDevitt,
  Influenza virus aerosols in human exhaled breath: particle size,
  culturability, and effect of surgical masks, PLoS pathogens 9~(3) (2013)
  e1003205.

\bibitem{leung2020}
N.~H. Leung, D.~K. Chu, E.~Y. Shiu, K.-H. Chan, J.~J. McDevitt, B.~J. Hau,
  H.-L. Yen, Y.~Li, D.~K. Ip, J.~M. Peiris, et~al., Respiratory virus shedding
  in exhaled breath and efficacy of face masks, Nature medicine 26~(5) (2020)
  676--680.

\bibitem{jacot2020}
D.~Jacot, G.~Greub, K.~Jaton, O.~Opota, Viral load of sars-cov-2 across
  patients and compared to other respiratory viruses, Microbes and infection
  22~(10) (2020) 617--621.

\bibitem{drivas1996}
P.~J. Drivas, P.~A. Valberg, B.~L. Murphy, R.~Wilson, Modeling indoor air
  exposure from short-term point source releases, Indoor Air 6~(4) (1996)
  271--277.

\bibitem{lee2003}
J.~H.-w. Lee, V.~Chu, V.~H. Chu, Turbulent jets and plumes: A Lagrangian
  approach, Vol.~1, Springer Science \& Business Media, 2003.

\bibitem{kukkonen1989}
J.~Kukkonen, T.~Vesala, M.~Kulmala, The interdependence of evaporation and
  settling for airborne freely falling droplets, Journal of aerosol science
  20~(7) (1989) 749--763.

\bibitem{vejerano2018}
E.~P. Vejerano, L.~C. Marr, Physico-chemical characteristics of evaporating
  respiratory fluid droplets, Journal of The Royal Society Interface 15~(139)
  (2018) 20170939.

\bibitem{petters2007}
M.~Petters, S.~Kreidenweis, A single parameter representation of hygroscopic
  growth and cloud condensation nucleus activity, Atmospheric Chemistry and
  Physics 7~(8) (2007) 1961--1971.

\bibitem{morris2020}
D.~H. Morris, K.~C.~H. Yinda, A.~Gamble, F.~W. Rossine, Q.~Huang, T.~Bushmaker,
  R.~J. Fischer, M.~J. Matson, N.~van Doremalen, P.~J. Vikesland, et~al., The
  effect of temperature and humidity on the stability of sars-cov-2 and other
  enveloped viruses, bioRxiv (2020).

\bibitem{van2020}
N.~Van~Doremalen, T.~Bushmaker, D.~H. Morris, M.~G. Holbrook, A.~Gamble, B.~N.
  Williamson, A.~Tamin, J.~L. Harcourt, N.~J. Thornburg, S.~I. Gerber, et~al.,
  Aerosol and surface stability of sars-cov-2 as compared with sars-cov-1, New
  England journal of medicine 382~(16) (2020) 1564--1567.

\bibitem{yang2012}
W.~Yang, S.~Elankumaran, L.~C. Marr, Relationship between humidity and
  influenza a viability in droplets and implications for influenza’s
  seasonality, PloS one 7~(10) (2012) e46789.

\bibitem{casanova2010}
L.~M. Casanova, S.~Jeon, W.~A. Rutala, D.~J. Weber, M.~D. Sobsey, Effects of
  air temperature and relative humidity on coronavirus survival on surfaces,
  Applied and environmental microbiology 76~(9) (2010) 2712--2717.

\bibitem{songer1967}
J.~R. Songer, Influence of relative humidity on the survival of some airborne
  viruses, Applied microbiology 15~(1) (1967) 35--42.

\bibitem{woese1960}
C.~Woese, Thermal inactivation of animal viruses, Annals of the New York
  Academy of Sciences 83~(4) (1960) 741--751.

\bibitem{polozov2008}
I.~V. Polozov, L.~Bezrukov, K.~Gawrisch, J.~Zimmerberg, Progressive ordering
  with decreasing temperature of the phospholipids of influenza virus, Nature
  chemical biology 4~(4) (2008) 248--255.

\bibitem{lin2019}
K.~Lin, L.~C. Marr, Humidity-dependent decay of viruses, but not bacteria, in
  aerosols and droplets follows disinfection kinetics, Environmental Science \&
  Technology 54~(2) (2019) 1024--1032.

\bibitem{kwon2020}
T.~Kwon, N.~N. Gaudreault, J.~A. Richt, Stability of sars-cov-2 on surfaces,
  bioRxiv (2020).

\bibitem{nguyen2014}
J.~L. Nguyen, J.~Schwartz, D.~W. Dockery, The relationship between indoor and
  outdoor temperature, apparent temperature, relative humidity, and absolute
  humidity, Indoor air 24~(1) (2014) 103--112.

\bibitem{cannell2006}
J.~Cannell, R.~Vieth, J.~Umhau, M.~Holick, W.~Grant, S.~Madronich, C.~Garland,
  E.~Giovannucci, Epidemic influenza and vitamin d, Epidemiology \& Infection
  134~(6) (2006) 1129--1140.

\bibitem{pieterse2018}
A.~Pieterse, S.~D. Hanekom, Criteria for enhancing mucus transport: a
  systematic scoping review, Multidisciplinary respiratory medicine 13~(1)
  (2018) 1--11.

\bibitem{earn2012}
D.~J. Earn, D.~He, M.~B. Loeb, K.~Fonseca, B.~E. Lee, J.~Dushoff, Effects of
  school closure on incidence of pandemic influenza in alberta, canada, Annals
  of internal medicine 156~(3) (2012) 173--181.

\bibitem{vouriot2021}
C.~V. Vouriot, H.~C. Burridge, C.~J. Noakes, P.~F. Linden, Seasonal variation
  in airborne infection risk in schools due to changes in ventilation inferred
  from monitored carbon dioxide, Indoor air (2021).

\bibitem{greenhalgh2021}
T.~Greenhalgh, J.~L. Jimenez, K.~A. Prather, Z.~Tufekci, D.~Fisman,
  R.~Schooley, Ten scientific reasons in support of airborne transmission of
  sars-cov-2, The Lancet (2021).

\end{thebibliography}





\end{document}